\documentclass{ws-procs9x6}

\begin{document}

\title{SYNTHESIS OF NEW NEUTRON RICH HEAVY NUCLEI:\\
AN EXPERIMENTALIST'S VIEW}
\author{W. LOVELAND$^*$ }

\address{Chemistry Department, Oregon State University,\\
Corvallis, Oregon 97331, USA\\
$^*$E-mail:lovelanw@onid.orst.edu}

\begin{abstract}
I attempt to experimentally evaluate the prospects of synthesizing new neutron-rich superheavy nuclei.  I consider three possible synthetic paths to neutron-rich superheavy nuclei: (a) the use of neutron-rich radioactive beams. (b) the use of damped collisions  and (c) the use of multi-nucleon transfer reactions.  I consider the prospects of synthesizing new n-rich isotopes of Rf-Bh using light n-rich radioactive beams and targeted beams  from ReA3, FRIB and SPIRAL2.  For the damped collision path, I present the results of a study of a surrogate reaction, $^{160}$Gd + $^{186}$W.  These data indicate the formation of Au (trans-target) fragments and the depletion of yields of target-like fragments by fission and fragment emission.  The data are compared to predictions of Zagrebaev and Greiner.  For the multi-nucleon transfer reactions, the results of a study of the $^{136}$Xe + $^{208}$Pb reaction are discussed.  I consider the possibility of multi-nucleon transfer reactions with radioactive beams.
\end{abstract}


\bodymatter

\section{Introduction}\label{aba:sec1}

In Figure 1, I show the current situation with regard to the synthesis of superheavy elements.  For cold fusion reactions one observes a steady decrease in the production cross section with increasing atomic number of the completely fused system.  The heaviest element reached using this reaction path is element 113, which is produced with a cross section of 22 $_{-13}^{+20}$ fb \cite{morita}.  Three events were observed in 553 days of beam time!!    For hot fusion reactions one observes a leveling out of the production cross sections around Ds (Z=110) with a slow decrease in cross sections up to element 118.  The recently reported  upper limits for the production of element 119  ( 55 fb)  and element 120 (160 fb) \cite{chris}  indicate a significant effort is required to proceed further.  The difficulty in proceeding towards the synthesis of heavier nuclei along with the promise of new opportunities in studying the more neutron rich isotopes of the heaviest elements has motivated increased efforts to make new neutron rich heavy nuclei.  The longer half-lives of the n-rich nuclei will enable more detailed atomic physics and chemistry studies.  For complete fusion reactions, the use of n-rich projectiles leads to lowered fusion barriers allowing production of nuclei at lower excitation energies with increased survival against fission.  For example, a comparison of the reactions of $^{32}$S and $^{38}$S with $^{208}$Pb at the nominal interaction barriers shows an enhanced cross section with the $^{38}$S projectile of a factor of 5000 along with the formation of a product whose half-life is 800 times longer.

\begin{figure}
\begin{center}
\psfig{file=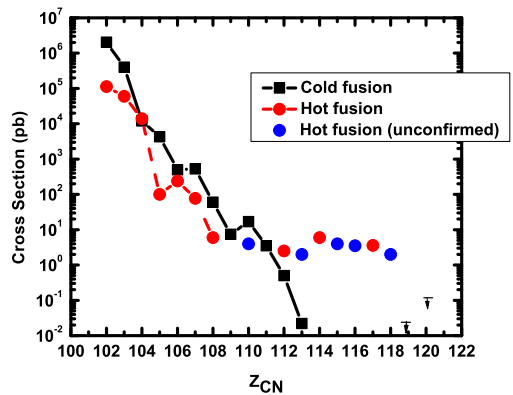,width=3.0in}
\end{center}
\caption{The cross sections for the production of superheavy nuclei as a function of the atomic number of the completely fused system.}
\label{aba:fig1}
\end{figure}

\section{Complete Fusion Reactions}

Most heavy element synthesis reactions to date have involved complete fusion reactions. For these reactions, the cross section for producing a heavy evaporation residue, $\sigma$$_{EVR}$, can be written as
\setcounter{equation}{0}
\begin{equation}
\sigma _{EVR}=\sum_{J=0}^{J_{\max }}\sigma
_{capture}(E_{c.m.},J)P_{CN}(E^{*},J) W_{sur}(E^{*},J)
\end{equation}
where $\sigma _{capture}(E_{c.m.},J)$ is the capture cross section at center of mass energy E$_{c.m.}$ and spin J. P$_{CN}$ is the probability that the projectile-target system will evolve from the contact configuration  inside the
fission saddle point to form a completely fused system rather than
re-separating (quasifission, fast fission). W$_{sur}$ is
the probability that the completely fused system will de-excite by neutron emission rather than fission. 

How well do we know these quantities?  Loveland \cite{wdl1} has evaluated the capture cross sections leading to heavy element synthesis and finds that these cross sections are known to within a factor of two.  Coupled channels calculations \cite{zaggy1} seem to do the best job of describing this quantity.  What about the survival probabilities, W$_{sur}$?  There is a well-established formalism to evaluate the survival probabilities \cite{vh}.  The principal uncertainty in these calculations is the values of the fission barrier heights used in the calculations.  The best calculations of the fission barrier heights for the heaviest nuclei show that average discrepancy between the measured and calculated values of the fission barrier heights is 0.4 MeV with the largest error being approximately 1 MeV. \cite{adam}  There is no indication of a systematic discrepancy between calculated and measured barrier heights as a function of isospin.  Based upon an evaluation of 75 heavy element synthesis reactions (where P$_{CN}$ = 1), Loveland \cite{wdl1}concluded that we know W$_{sur}$ within a factor of 3. 

 The least well known quantity is P$_{CN}$ as it is difficult to measure and to calculate.  Zagrebaev and Greiner \cite{zaggy2} have suggested the following functional form for the excitation energy dependence of P$_{CN}$
\begin{equation}
P_{CN}(E^{\ast },J)=\frac{P_{CN}^{0}}{1+\exp \left[ \frac{E_{B}^{\ast
}-E_{int}^{\ast }(J)}{\Delta }\right] }
\end{equation}
where P$_{CN}^{0}$ is the asymptotic fissility dependent value of P$_{CN}$, E$_{B}^{\ast}$ is the excitation energy at the Bass barrier, E$_{int}^{\ast }$(J) is the internal excitation energy (E$_{c.m.}$+Q-E$_{rot}$(J)) and $\Delta$ is 4 MeV.  A comparison of this formula with data \cite {34, wdl1} shows an excellent agreement between data and predictions.  Loveland \cite{wdl1} considered a group of measurements of P$_{CN}$ for hot fusion reactions where E* was 40-50 MeV and found a satisfactory empirical description of the dependence of P$_{CN}$ upon fissility.  All in all, P$_{CN}$ is probably known to within an order of magnitude.

\section{Complete Fusion with Radioactive Beams}

We can apply what we know about the synthesis of the heaviest nuclei to the problem of making new heavy nuclei with radioactive beams.  The calculational model I employ \cite{wdl} is simple and unsophisticated.  One takes the beam list for any radioactive beam facility (FRIB, SPIRAL2, ReA3, CARIBU, etc.\cite{67,68,69,70,71,72}) and then considers every possible combination of a radioactive projectile with all ``stable" targets.  One varies the projectile energy and evaluates $\sigma$$_{capture}$, P$_{CN}$ and W$_{sur}$ to get $\sigma$$_{EVR}$.  From this, one uses reasonable assumptions about target thickness (0.5 mg/cm$^{2}$) and calculates the product yield in atoms/day.  

The central issue in using radioactive beams to synthesize heavy nuclei is the intensity of the radioactive beams.  Stable beams are routinely available at intensities of 6 x 10$^{12}$ /s and current plans are to increase these intensities by 1-2 orders of magnitude.  Radioactive beams are rarely available at intensities of 10$^{10}$ - 10$^{11}$ /s with substantial decreases in intensity for more n-rich beams.  As a consequence, despite the enhanced production cross sections with radioactive beams, the best production rates of heavy nuclei are 3 orders of magnitude less than those achieved stable beams.  The consequence of this is that  { \bf radioactive beams are not a pathway to new superheavy elements.}  Does that mean that radioactive beams are worthless when it comes to making new heavy nuclei?  No, radioactive beams are useful tools for producing new n-rich isotopes of elements 104-107.  In table 1  I show a list of new n-rich isotopes of elements 104-107 that can be made at rates greater than 5 atoms/day and the reactions that produce them.

\begin{table}
\tbl{Reactions predicted to form 5 or more atoms per day of new n-rich nuclei.} 
{\begin{tabular}{@{}cc@{}}
\toprule
Nucleus&Reaction\\
$^{264}$Rf&$^{252}$Cf($^{16}$C,4n)\\
$^{265}$Db&$^{249}$Bk($^{20}$O,4n)\\
$^{268}$Sg&$^{252}$Cf($^{20}$O,4n)\\
$^{267}$Bh &$^{252}$Cf($^{21}$F,6n)\\
\botrule
\end{tabular}
}
\label{aba:tbl1}
\end{table}

One might pose the question as to which radioactive beams are projected to be the most useful in synthesizing these n-rich nuclei.  The answer to this question is the light beams such as O, Ne, Mg, etc. because of their high intensities.  In table 2 I show the reactions and rates for the production of  n-rich isotopes of Sg, which all involve these light nuclei.

\begin{table}
\tbl{Typical reactions that form  n-rich isotopes of Sg.}
{\begin{tabular}{@{}cccc@{}}
\toprule
Reactants&Products &FRIB beam intensity (p/s)&Production Rate (atoms/day)\\
$^{26}$Ne + $^{248}$Cm&$^{271}$Sg + 3n&2.2 x 10$^{6}$&0.004\\
$^{30}$Mg + $^{244}$Pu&$^{270}$Sg + 4n&7.1 x 10$^{6}$&1\\
$^{29}$Mg + $^{244}$Pu&$^{269}$Sg + 4n&3.6 x 10$^{7}$&0.2\\
$^{20}$O + $^{252}$Cf&$^{268}$Sg + 4n&1.5 x 10$^{8}$&5\\
$^{23}$Ne + $^{248}$Cm&$^{267}$Sg + 4n&1.6 x 10$^{8}$&1\\
\botrule
\end{tabular}
}
\label{aba:tbl2}
\end{table}

Please note though that $^{271}$Sg is on the n-poor side of beta stability and beta stability for Sg is at 276.

A recent related issue is that of targeted radioactive beams.  It may be possible, by special efforts, to produce beams of the K and/or Ar isotopes that would be useful in heavy element synthesis.  For example, $^{46}$At, produced by $^{48}$Ca fragmentation, could be used to synthesize $^{286,287}$Cn at atom per day rates if this beam was available at 10$^{10}$ ions/s.

\section{Damped Collisions}
Recently , there has
been a revival of interest in the use of damped collisions of massive nuclei
at near barrier energies to synthesize superheavy nuclei, particularly
those nuclei with large neutron excess, approaching the N=184 shell. In the
1980s \cite{olddata} there were attempts to use the $^{238}$U + $^{238}$U and
the $^{238}$U + $^{248}$Cm reactions at above barrier energies to produce trans-target nuclides. While
there was evidence for the formation of neutron-rich isotopes of Fm and Md
at the 0.1 $\mu $b level, no higher actinides were found. The fundamental
problem was that the nuclei that were produced far above the target nucleus
were the result of events with high total kinetic energy loss, i.e., high
excitation energies and resulting poor survival probabilities. Very
recently, Zagrebaev and Greiner \cite{51,52,53,54,55,56,57,58} using a new
model \cite{59} for these collisions, have examined the older
experiments and some proposed new experiments ($^{232}$Th +$^{250}$Cf, $%
^{238}$U+$^{238}$U, and $^{238}$U +$^{248}$Cm). With their new model which
emphasizes the role of shell effects in damped collisions, they are able to
correctly describe the previously measured fragment angular, energy and
charge distributions from the $^{136}$Xe + $^{209}$Bi reaction and the
isotopic yields of Cf , Es, Fm and Md from the $^{238}$U + $^{248}$Cm
reaction. They predict that by a careful choice of beam energies and
projectile-target combinations, one might be able to produce n-rich isotopes
of element 112 in the $^{248}$Cm +$^{250}$Cf reaction. They suggest the
detection of $^{267,268}$Db and $^{272,271}$Bh (at the pb level) in the Th +
Cf or U + Cm reactions to verify these predictions. Such experiments are
very difficult because of the low cross sections, the lower intensities of
these massive projectile beams and the problems of detecting the reaction
products in an ocean of elastically scattered particles, etc.

However, in 2007, Zagrebaev and Greiner \cite{63} outlined a simpler
test of their theoretical predictions. They applied the same model used to
study the U + Cm, Th + Cf and U + U collisions to the $^{160}$Gd + $^{186}$W
reaction. As an experimentalist, I really appreciate this suggestion of a surrogate reaction that allows one to check the theoretical predictions in a more accessible  system. 

In figure [2] I show the results of an experimental study using radiochemical techniques of the $^{160}$Gd + $^{186}$W reaction.\cite{wdl2}  Both the measured and predicted mass distributions show the expected ``rabbit ears". i.e, a peak in the yields near the mass of the target and the projectile nuclei.  The measured distribution shows yields of what are probably fission fragments and intermediate mass fragments not predicted by the model.  Perhaps the most significant feature of the mass distribution is a peak in the mass distribution for trans target nuclei (A= 190-200)   This trans target peak appears to be at Z=79 (Au), reminding one of the ``goldfinger" seen in studies of low energy deep inelastic scattering in the 1970s.  All this is consistent of the formation of a much heavier product that decays by fission and then particle emission to give rise to this trans target peak.  Zagrebaev and Greiner had actually predicted enhanced trans target yields in the Pb isotopes, which were searched for but not observed.  This result and the results of the TAMU group \cite{joe1,joe2}for the 7.5 A MeV $^{197}$Au + $^{232}$Th reaction are encouraging for the effort to use these reactions to produce new n-rich heavy nuclei.

\begin{figure}
\begin{center}
\psfig{file= 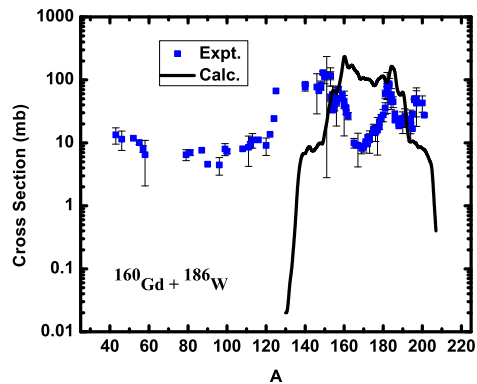, width=3.0in}
\end{center}
\caption{Comparison of measured \cite{wdl2} and predicted \cite{63} values of the fragment mass distribution for the reaction of $^{160}$Gd with $^{186}$W.}
\label{aba:fig2}
\end{figure}

\section{Multi-nucleon Transfer}

Zagrebaev and Greiner have suggested that multi-nucleon transfer reactions might be an alternate path to n-rich heavy nuclei \cite{15}.  In the reaction of $^{238}$U with $^{248}$Cm, they predict the formation of $^{262,264,266}$No with cross sections ranging from 10 to 1000 pb and the formation of $^{270}$Sg with a cross section of $\sim$ 20 pb.  (With current $^{238}$U beam intensities, this is a production rate of $\sim$ 1 atom/day, similar to that predicted with radioactive beams.)  They have also provided us with an interesting surrogate for these reactions, the reaction of $^{136}$Xe with $^{208}$Pb, where tens of new n-rich nuclei could be produced with cross sections higher than 1 $\mu$b.  The first new published study of these possibilities \cite{koz}, similar to those seen in damped collisions \cite{wdl2}, showed the formation of n-rich Rn and Ra nuclei with unexpectedly large cross sections.

Perhaps the most interesting possibility is the use of radioactive beams in multi-nucleon transfer reactions.  For example Pollarollo \cite{xx} and Zagrebaev \cite {yy} have predicted the formation of the very n-rich isotopes (N=152-162) of U-Cm in the reaction of  950 MeV $^{144}$Xe + $^{248}$Cm with $\mu$b or greater cross sections.

\section{Conclusions}
Radioactive beams and damped collisions may offer us the opportunity of making more n-rich heavy nuclei although there seems to be no clear path to making nuclei on the n-rich side of stability.  Multi-nucleon transfer reactions appear to allow us the opportunity of reaching nuclei on the n-rich side of stability with atomic numbers below that of the targets used in these reactions.

\section{Acknowledgments}
This work was supported in part by the Office of High Energy and Nuclear Physics, Nuclear Physics Division, US Department of Energy, under Grant No. DE-FG06-97ER41026

\bibliographystyle{ws-procs9x6}
\bibliography{ws-pro-sample}

\begin{thebibliography}{9}

\bibitem{morita} K. Morita, et al.,{\em J. Phys. Soc. Jpn.} {\bf 81}, 103201 (2012).
\bibitem{chris} C. H. Duellmann, private communication, 2012
\bibitem{wdl1}W. Loveland, J. Phys. Conf. Series, accepted for publication
\bibitem{zaggy1}http://nrv.jinr.ru/nrv/webnrv/fusion/
\bibitem{67}https://groups.nscl.msu.edu/frib/rates/fribrates.html
\bibitem{68}http://www.nscl.msu.edu/exp/sr/yields
\bibitem{69}http://pro.ganil-spiral2.eu/spiral2/spiral2-beams/radioactive-ion-beams-of-spiral2/post-accelerated-cime-isol-rib-beams-available-for-the-day-1-spiral2-phase-2-experiments
\bibitem{70}http://pro.ganil-spiral2.eu/spiral2/spiral2-beams/radioactive-ion-beams-of-spiral2/radioactive-ion-beams-of-spiral2-of-fission-fragments/view
\bibitem{71}http://www.phy.anl.gov/atlas/facility/caribubeams.html
\bibitem{72}C.L. Jiang,B.B. Back, I. Gomes,A.M. Heinz, J. Nolen, K.E.  Rehm, G. Savard and J.P.Schiffer,  {\em Nucl. Instru. Meth. Phys. Res. A} {\bf 492} 57 (2002).
\bibitem{vh} R. Vandenbosch  and J.R.  Huizenga,  Nuclear Fission (Academic, New York, 1973), p. 323
\bibitem{adam}M. Kowal, P. Jachimowicz,  and A.  Sobiczewski,  {\em Phys. Rev. C} {\bf 82} , 014303(2010).
\bibitem{zaggy2} V. Zagrebaev  and W. Greiner  {\em Phys. Rev. C} {\bf 78} 034610 (2008).
\bibitem{34} G.N. Knyazheva , et al. {\em  Phys. Rev.C} {\bf 75}  064602 (2007).
\bibitem{wdl} W. Loveland {\em Phys. Rev.} C {\bf 78}, 014612 (2007).
\bibitem{olddata}See G.T. Seaborg and W. Loveland, The Elements Beyond Uranium (Wiley, New York, 1990) for a review of these data.
\bibitem{51} V.I. Zagrebaev, Y.T. Oganessian, M.G. Itkis  and W. Greiner {\em Phys. Rev. C} {\bf 73}  031602(R) (2006).
\bibitem{52}V. Zagrebaev  and W. Greiner   {\em J. Phys. G} { \bf 34} 1 (2007).
\bibitem{53} V. Zagrebaev  and W. Greiner    {\em J. Phys. G} { \bf 35} 125103 (2008).
\bibitem{54}V. Zagrebaev  and W. Greiner   {\em Phys. Rev. Lett.} {\bf 101} 122701(2008).
\bibitem{55}V. Zagrebaev  and W. Greiner  CP1098, FUSION08: New Aspects of
Heavy Ion Collisions Near the Coulomb Barrier, K.E. Rehm, B.B. Back, H.
Esbensen, and C.J. Lister, ed., (AIP, New York, 2009) pp326-333.
\bibitem{56}V. Zagrebaev   and W. Greiner   {\em Nucl. Phys. A} {\bf834} 366c (2010).
\bibitem{57}V. Zagrebaev  and W. Greiner  {\em Phys. Rev.  C} {\bf78}  034610 (2008).
\bibitem{58}V. Zagrebaev  and W. Greiner  {\em Russ. Chem. Rev.} { \bf78} 1089 (2009).
\bibitem{59}V. Zagrebaev and W. Greiner  {\em J. Phys. G} { \bf 31}  825 (2005).
\bibitem{63}V. Zagrebaev  and W. Greiner  {\em J. Phys. G} {\bf 34}  2265 (2007).
\bibitem{wdl2} W. Loveland, A.M. Vinodkumar, D. Peterson,  and J.P. Greene   {\em Phys. Rev. C} {\bf 83} 044610 (2011).
\bibitem{joe1} M. Barbui  et al.   {\em Intl. J. Mod. Phys. E} {\bf 18} 1036 (2009).
\bibitem{joe2}  M. Barbui  et al.   {\em J. Phys.: Conf. Series} {\bf 312}  082012 (2011).
\bibitem{15} V.I. Zagrebaev and W. Greiner, {\em Phys. Rev. C} {\bf 83}, 044618 (2011).
\bibitem{koz} E. M. Kozulin et al., {\em Phys. Rev C} {\bf 86}, 044611 (2012).
\bibitem{xx} G. Pollarollo (private communication)
\bibitem{yy} V. Zagrebaev (private communication).

\end{thebibliography}

\end{document}